\documentclass[journal = jpcc, manuscript = article]{achemso}
\usepackage{setspace}
\setkeys{acs}{articletitle = true, etalmode=truncate, maxauthors = 10}
\usepackage{graphicx}
\usepackage{amsmath,amssymb,amsfonts,bm}
\usepackage{mathtools}
\usepackage{braket}

\author{Amos Afugu}
\affiliation{Department of Chemistry, Wayne State University, Detroit, Michigan 48202, USA}
\author{Gyanu P. Kafle}
\affiliation{Department of Chemistry, Wayne State University, Detroit, Michigan 48202, USA}
\author{Zhen-Fei Liu}
\email{zfliu@wayne.edu}
\affiliation{Department of Chemistry, Wayne State University, Detroit, Michigan 48202, USA}
\title{First-Principles Analysis of Chirality-Induced Spin Selectivity at Molecule-Metal Interfaces in Photoemission}

\begin{document}

\begin{abstract}
Spin-resolved photoelectron spectroscopy (PES) probes chirality-induced spin selectivity (CISS), yet it remains unclear whether the measured spin polarization reflects molecular chirality itself or the broader electronic structure of the hybrid interface. We present a first-principles analysis of PES spin polarization at chiral molecule-metal interfaces, treating the interface holistically rather than as a metal substrate plus a separate molecular spin filter/polarizer. Using density functional theory within a three-step photoemission framework, we compute the spin polarization generated in the optical-excitation step for ($M$)- and ($P$)-heptahelicene adsorbed on Au(111) and Cu(111), and for coronene/Au(111) as a non-chiral control. We find that adsorption strongly reshapes the PES spin polarization relative to the clean metal surface, but opposite enantiomers yield symmetry-related responses. These results indicate that changes in the PES spin polarization are more naturally attributed to the electronic structure of the hybrid interface than to molecular chirality alone.

\end{abstract}

Chirality-induced spin selectivity (CISS) is an emerging field that studies the coupling between chirality -- a fundamental geometric property of molecules -- and electron spin, an intrinsic quantum degree of freedom.\cite{naaman2019,bloom24} The term CISS encompasses a broad range of phenomena, including those observed in photoelectron spectroscopy (PES),\cite{Ray1999,mollers2022b} electron transport and tunneling,\cite{xie2011,albro2025} and catalysis.\cite{mtangi2015,mondal2015} Among these, spin-resolved PES is often regarded as the gold standard for characterizing the CISS effect\cite{bloom24} and has been applied to both helical and point-chiral systems, such as DNA,\cite{Gohler2011} proteins,\cite{mishra2013} heptahelicene,\cite{Kettner2018,baljozovic2023} helical tetrapyrrole,\cite{mollers2024} methylcyclohexanone,\cite{chetana2022} and chiral CuO thin film.\cite{mollers2022} In these experiments, chiral molecules are adsorbed on a metal surface, and the spin polarization of photoelectrons emitted from the metal surface (more precisely, from the molecule-metal interface) is measured and compared to that of the clean metal surface without molecular adsorbates. Reported observations have often been interpreted in terms of a handedness-dependent change in photoelectron spin polarization, in some cases even on substrates with weak intrinsic spin-orbit coupling (SOC).\cite{mishra2013} These findings call for a rigorous microscopic understanding of the origin of CISS effects reported in PES measurements, as well as a careful and critical assessment of the respective roles of the chiral molecule and the hybrid molecule-metal interface.

A variety of theoretical models have been advanced to rationalize these experimental findings,\cite{evers2022} including approaches based on scattering theory,\cite{yeganeh2009,gersten2013,gutierrez2013,michaeli2019,geyer2020} symmetry considerations,\cite{rashba2003,dalum2019} electron correlation,\cite{fransson2021,chiesa2024,xu2024} nonequilibrium dynamics,\cite{guo2012,zhang2025,smorka2025} and spinterface physics.\cite{monti2024} Despite these efforts, a comprehensive first-principles computational framework remains lacking. With the SOC treated at the \emph{ab initio} level, the computed spin polarization is typically orders of magnitude smaller than experiments.\cite{zollner2020,zollner-b-2020} Notably, most prior first-principles studies have focused on CISS in electron transport below the vacuum level,\cite{naskar2023,day2024} where the operative mechanisms differ fundamentally from those governing PES, which probes excited states above the vacuum level and involves distinct initial and final states. Consequently, methods developed to explain CISS in transport measurements may not be directly transferable to PES. Furthermore, many existing treatments conceptually separate the chiral molecular adsorbate from the metal substrate, emphasizing either spin-filtering\cite{gutierrez2012,gersten2013} or spin-polarizing\cite{wolf2022} effects within the molecular layer on the electron flow originating from the metal. Such a viewpoint overlooks the central role of the molecule-metal interface, neglecting both the adsorbate-induced modification of the metal electronic structure and emergent interfacial states arising from the molecule-metal hybridization.

In this work, we treat the molecule-metal interface holistically as the relevant microscopic object, including SOC at the first-principles level. Within this picture, the change in photoelectron spin polarization is not attributed solely to the molecular layer, but rather emerges from the hybridized interface, where adsorbate-induced modulation of the metal states gives rise to spin-dependent photoemission different from that of the clean metal surface. We believe this viewpoint highlights interfacial states as the central mediators that naturally connect the intrinsic large SOC in the metal to the spin polarization in PES. We compute PES intensities using a three-step model\cite{Hufner2003PES,Damascelli2003} and focus our discussion on the first step, i.e., the optical excitation that involves initial and final states of the interface, as well as their spin texture and the polarization of the incident light. We isolate this step to identify qualitative trends, and defer the effects associated with the other two steps (transport and escape) and a more realistic final-state description to future work. For conceptual clarity, we employ an independent-particle description in which both the initial and final states are approximated by Kohn-Sham spinors computed from density functional theory (DFT). This approximation does not include all many-body and final-state effects, but it provides a controlled first-principles framework for identifying which features of the PES spin polarization are already encoded in the hybridized interface. Later in the paper, we also assess the effect of many-body corrections on the interfacial level alignment.\cite{liu2025}

We define the spin polarization along the $z$ direction as:
\begin{equation}
\zeta_z\left(\mathbf{K}_\parallel\right) = \frac{I_{\mathbf{K}_\parallel}\left(\ket{S_z\,\uparrow}\right)-I_{\mathbf{K}_\parallel}\left(\ket{S_z\,\downarrow}\right)}{I_{\mathbf{K}_\parallel}\left(\ket{S_z\,\uparrow}\right)+I_{\mathbf{K}_\parallel}\left(\ket{S_z\,\downarrow}\right)}\times 100\%.
\label{eq:zetaz}
\end{equation}
In this equation, $\mathbf{K}_\parallel$ denotes the in-plane (parallel to the metal surface) momentum of the photoelectrons. $I_{\mathbf{K}_\parallel}\left(\ket{S_z\,\uparrow}\right)$ denotes the photoemission intensity for detecting photoelectrons with in-plane momentum $\mathbf{K}_\parallel$ in the eigenstate $\ket{S_z\,\uparrow}$ of $\hat{S}_z$ with eigenvalue $\hbar/2$. Generalizing Fermi's golden rule to a degenerate final-state manifold, as detailed in the Supporting Information, we can write

\begin{equation}
\begin{split}
I_{\mathbf{K}_\parallel}\left(\ket{S_z\,\uparrow}\right) \propto & \sum_{\rm i} \sum_{\rm f'}\left\Vert \braket{S_z\,\uparrow|\phi_{{\rm i}\to\rm f'}}\right\Vert^2 \delta\left(\epsilon_{\rm f'}-\epsilon_{\rm i}-h\nu\right) \\
& \times\delta\left(E_{\rm kin}-\left[\epsilon_{\rm f'}-E_{\rm vac}\right]\right)\delta\left(\mathbf{K}_\parallel-\mathbf{k}_\parallel-\mathbf{G}_\parallel\right).
\end{split}
\label{eq:isz}
\end{equation}
Here, because the goal is to compute the unitless spin polarization in Eq. \eqref{eq:zetaz}, we have neglected the scalar prefactor in Eq. \eqref{eq:isz} that is identical for both spin channels. $\ket{\phi_{{\rm i}\to\rm f'}}$ is an optically excited spinor defined in the Supporting Information, and $\left\Vert \braket{S_z\,\uparrow|\phi_{{\rm i}\to\rm f'}}\right\Vert^2$ represents the squared norm of the spatial component of the excited spinor after projection onto $\ket{S_z\,\uparrow}$. We emphasize that in this equation, transition amplitudes into states within the same final-state degenerate manifold are summed coherently,\cite{wang2011,park2012} while contributions from different occupied initial states and distinct final-state degenerate manifolds need to be summed at the probability level (see Supporting Information for details).

The first delta function in Eq. \eqref{eq:isz} enforces energy conservation during the optical excitation from an initial state with energy $\epsilon_{\rm i}$ to a degenerate final-state manifold with energy $\epsilon_{\rm f'}$, and $h\nu$ is the photon energy. The second delta function enforces energy conservation for photoelectrons escaping into vacuum, where $E_{\rm kin}$ is the kinetic energy of the emitted photoelectron and $E_{\rm vac}$ is the vacuum level.\cite{Hufner2003PES,Damascelli2003} The last delta function enforces conservation of in-plane momentum,\cite{Hufner2003PES} where $\mathbf{k}_\parallel$ and $\mathbf{G}_\parallel$ are the in-plane crystal momentum and reciprocal-space lattice vector in the Bloch wavefunctions. Technically, this is implemented such that we truncate the spinor $\ket{\phi_{{\rm i}\to\rm f'}}=\ket{\phi_{{\rm i}\to\rm f'}^\alpha}\ket{S_z\,\uparrow}+\ket{\phi_{{\rm i}\to\rm f'}^\beta}\ket{S_z\,\downarrow}$ to certain plane-wave components:
\begin{equation}
\begin{split}
\ket{\phi_{{\rm i}\to\rm f'}^\alpha}& = e^{{\rm i}\mathbf{k}_\parallel \cdot \mathbf{r}} \sum_\mathbf{G} c_\mathbf{G}^\alpha e^{{\rm i}\mathbf{G} \cdot \mathbf{r}} \\
& \to e^{{\rm i}\mathbf{k}_\parallel \cdot \mathbf{r}} \sum_{G_z>0}\sum_{\mathbf{G}_\parallel = \mathbf{K}_\parallel - \mathbf{k}_\parallel} c_\mathbf{G}^\alpha e^{{\rm i}\mathbf{G} \cdot \mathbf{r}}.
\end{split}
\label{eq:pwtrunc}
\end{equation}
This is written for $\ket{\phi_{{\rm i}\to\rm f'}^\alpha}$ and a similar treatment for $\ket{\phi_{{\rm i}\to\rm f'}^\beta}$ applies. In the second line of Eq. \eqref{eq:pwtrunc}, we restrict the summation to $G_z>0$ because only those plane-wave components can escape from the surface. In this work, we only consider $\mathbf{K}_\parallel$ values that match our $\mathbf{k}_\parallel$ sampling in the calculation, such that the second summation can be restricted to $\mathbf{G}_\parallel=0$ in these special cases. 

Analogous definitions apply for $\zeta_x$ and $\zeta_y$, or more generally for spin polarization along an arbitrary quantization axis. Equivalently, Eq. \eqref{eq:zetaz} can be expressed in density-matrix form as $\mbox{Tr}\left(\rho\sigma_z\right)/\mbox{Tr}\rho$, where $\sigma_z$ is a Pauli matrix and $\rho=\sum_{\rm {i}}\sum_{\rm f'}\ket{\phi_{{\rm i}\to\rm f'}}\bra{\phi_{{\rm i}\to\rm f'}}$ is the spin density matrix of the photoexcited states. This formulation has been employed in the analysis of the spin polarization of photoelectrons from nonmagnetic solids and surfaces, including Au.\cite{borstel1982,borstel2-1982,wohlecke1984}

\begin{figure}[!ht]
\centering
\includegraphics[width=3.3in]{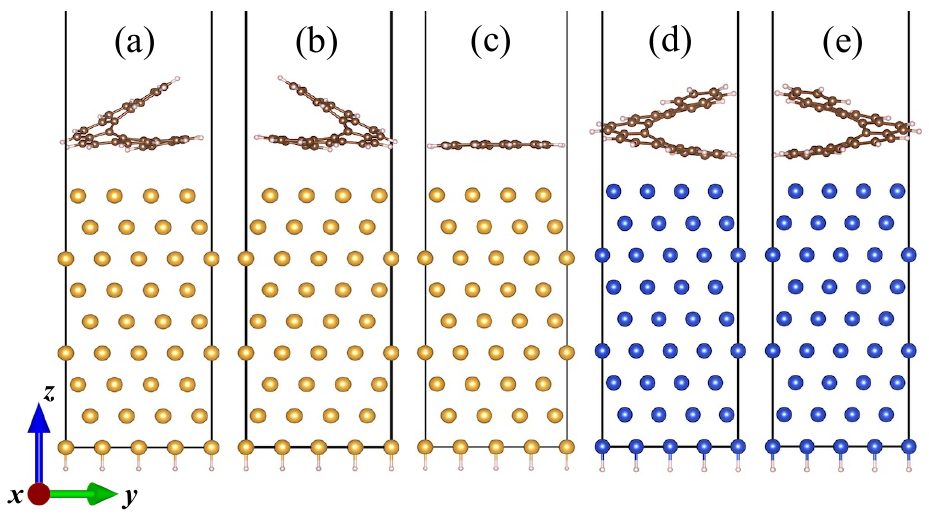}
\caption{Interfaces studied in this work. (a) ($M$)-[7]H, (b) ($P$)-[7]H, and (c) coronene adsorbed on Au (111). (d) ($M$)-[7]H and (e) ($P$)-[7]H adsorbed on Cu (111). The bottom metal surface is passivated with hydrogen atoms.}
\label{fig:stru}
\end{figure}

In this work, we compute both in-plane (parallel to the metal surface in Figure \ref{fig:stru}) spin polarizations $\zeta_x^\pm$ and $\zeta_y^\pm$, and the out-of-plane (along the surface normal in Figure \ref{fig:stru}) component, $\zeta_z^\pm$, the latter being the quantity most commonly reported in experimental studies.\cite{Kettner2018,Gohler2011} The superscript ``$\pm$'' corresponds to the helicities of circularly polarized light (see Supporting Information). We also compute $\zeta_z^{\rm LP}$, the out-of-plane spin polarization resulting from linearly polarized incident light. As follows from Eq. \eqref{eq:isz}, all spin polarization components depend explicitly on the kinetic energy $E_{\rm kin}$ of the emitted photoelectrons. We therefore present $\zeta(E_{\rm kin})$ throughout, noting that the maximum accessible $E_{\rm kin}$ is constrained by energy conservation to $h\nu$ minus the surface work function. We evaluate the spin polarization both at the $\Gamma$ point of the surface Brillouin zone and at finite $\mathbf{k}_\parallel$ points away from $\Gamma$. The former corresponds to normal-emission PES measurements commonly performed in the literature,\cite{Kettner2018,Gohler2011} although some experiments do report $\mathbf{k}_\parallel$-resolved spin polarization.\cite{baljozovic2023} In all calculations, the first two delta functions in Eq. \eqref{eq:isz} for energy conservation are approximated by Lorentzians: $\delta(x_1-x_2)\approx\gamma/[(x_1-x_2)^2+\gamma^2]$ with $\gamma=0.01$ eV.

We apply this theoretical framework to investigate the spin polarization of photoelectrons emitted from a series of molecule-metal interfaces shown in Figure \ref{fig:stru}. Specifically, we consider the chiral molecules ($M$)-heptahelicene ($M$-[7]H, C$_{30}$H$_{18}$, $M$ denoting minus helicity) and ($P$)-[7]H ($P$ denoting plus helicity) adsorbed on Au (111), as shown in Figure \ref{fig:stru}(a) and (b), respectively. These systems correspond to those examined experimentally in Ref. \citenum{Kettner2018}. As a non-chiral control, we additionally study coronene (C$_{24}$H$_{12}$) adsorbed on Au (111), shown in Figure \ref{fig:stru}(c). We choose this system due to its comparable lateral size and conjugation length to [7]H: it consists of seven peri-fused benzene rings, analogous to the seven annulated benzene units in [7]H, but lacks intrinsic chirality. To further assess the role of substrate SOC, we also investigate ($M$)-[7]H and ($P$)-[7]H adsorbed on Cu (111), as shown in Figure \ref{fig:stru}(d) and (e), respectively. These systems were studied experimentally in Ref. \citenum{baljozovic2023}. This combination of systems enables us to disentangle three effects: the intrinsic response of the clean substrate, the modification induced by adsorption of molecules, and the portion of that modification that is genuinely specific to molecular chirality. We primarily use $h\nu=$5.83 eV as the energy of the incident light in the calculations (consistent with experiment\cite{Kettner2018,baljozovic2023}) and also report results using other photon energies in the Supporting Information.

Details of the geometry optimization and other considerations of the modeling are discussed in the Supporting Information. We explicitly relax the ($M$)-[7]H/Au, coronene/Au, and ($M$)-[7]H/Cu interfaces, while the corresponding ($P$)-[7]H interface structures are generated by taking the mirror image of the ($M$)-[7]H configurations with respect to the $xz$ plane. This ensures that any differences observed between the two enantiomers arise from electronic effects rather than from  geometry variations.


\begin{figure}[!ht]
\centering
\includegraphics[width=3.3in]{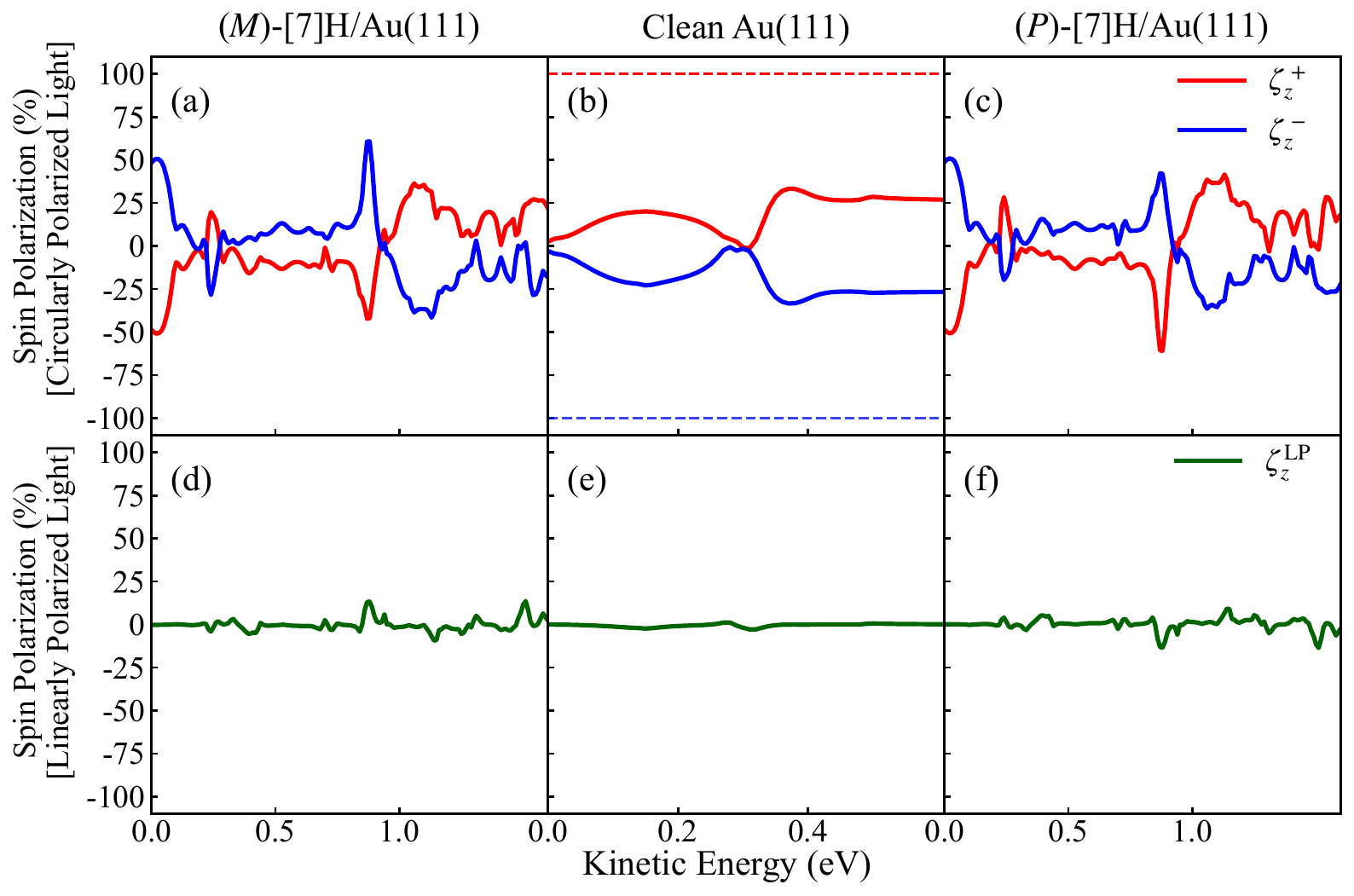}
\caption{Spin polarization along the $z$ direction at the $\Gamma$ point. (a)(d) ($M$)-[7]H adsorbed on Au (111). (b)(e) Clean Au (111), where dashed lines are computed for a nine-layer unit cell and solid lines are computed for an Au slab with coordinates adopted from the ($M$)-[7]H/Au interface. (c)(f) ($P$)-[7]H adsorbed on Au (111). (a-c) Incident light is circularly polarized. Red (blue) lines denote $\zeta^+$ ($\zeta^-$), spin polarization under right (left) circularly polarized light. (d-f) Incident light is linearly polarized.}
\label{fig:Au}
\end{figure}

We begin with the clean Au (111) surface in the absence of molecular adsorbates. Figure \ref{fig:Au}(b) presents the out-of-plane spin polarization $\zeta_z^\pm(E_{\rm kin})$ at the $\Gamma$ point, under right (``$+$'') and left (``$-$'') circularly polarized light with photon energy $h\nu=$ 5.83 eV. Dashed lines denote results obtained using a nine-layer Au (111) unit cell, whereas solid lines correspond to a $4\times 4\times 9$ Au (111) slab with the atomic coordinates adopted from the relaxed ($M$)-[7]H/Au interface system. The difference between dashed and solid curves arises exclusively from geometry relaxation of the molecule-metal interface, as calculations performed for a fully periodic $4\times 4\times 9$ Au (111) supercell without molecular adsorbates simply reproduce the results of the nine-layer Au (111) unit cell. 


The spin polarization of the Au (111) unit cell reflects both the symmetry of the surface and the specific photon energy used. With $h\nu=$ 5.83 eV and at the $\Gamma$ point, we find $\zeta_z^\pm=\pm 1$ and both in-plane components $\zeta_x^\pm$ and $\zeta_y^\pm$ vanish by symmetry. Figure S2 of the Supporting Information summarizes the behaviors of $\zeta_x^\pm$, $\zeta_y^\pm$, and $\zeta_z^\pm$ at four additional representative $\mathbf{k}_\parallel$ points in the vicinity of $\Gamma$. For $h\nu=$ 5.83 eV, the photoemission process involves only the Shockley surface states located approximately 0.5 eV below the Fermi level, which are doubly degenerate at $\Gamma$ (Kramers degeneracy) and spin split at all other $\mathbf{k}_\parallel$ points. When a larger photon energy is used, additional initial states contribute. As shown in Figure S3 of the Supporting Information, for $h\nu=$ 8 eV and 10 eV, $\zeta_z^\pm$ at $\Gamma$ is reduced below unity, which we attribute to destructive interference among transitions originating from multiple initial states. 


This idealized behavior is strongly reduced once the Au geometry is taken from the relaxed molecule-metal interface, even after removing the molecule. Figure \ref{fig:Au}(b) presents the results for the $4\times 4\times 9$ Au (111) slab whose atomic coordinates are adopted from the relaxed ($M$)-[7]H/Au interface, where we observe qualitative changes in the spin polarization compared to the primitive Au (111) unit cell. The calculated work function of this slab is 5.20 eV, implying that the incident light with $h\nu=$ 5.83 eV can probe electronic states up to 0.63 eV below the Fermi level. At the $\Gamma$ point, the out-of-plane spin polarization $\zeta_z^\pm$ decreases from unity to about 20\%, with $\zeta_z^+= -\zeta_z^-$ for all $E_{\rm kin}$. This change can be understood as a consequence of symmetry lowering and the associated mixing of irreducible representations in the initial and/or final states.\cite{wohlecke1984} The comparison therefore suggests that even modest departures from perfect surface symmetry can strongly alter the computed PES spin polarization. This needs to be taken into account when connecting theory to experiment. Rather than taking the ideal $\zeta_z^\pm=\pm1$ result as the relevant experimental baseline, the lesson is that the spin polarization of Au(111) is highly sensitive to the actual local geometry of the interface. Small random distortions on top of a perfect Au (111) slab lead to the same qualitative conclusion. Figure \ref{fig:Au}(e) presents the corresponding result for linearly polarized incident light, $\zeta_z^{\rm LP}$, which is nearly zero for all $E_{\rm kin}$, consistent with Refs. \citenum{Kettner2018} and \citenum{Hoesch2004}.

We next consider ($M$)-[7]H and ($P$)-[7]H adsorbed on Au (111). Molecular adsorption lowers the work function to 4.22 eV. Consequently, for incident light with $h\nu=$ 5.83 eV, the maximum kinetic energy of the emitted electrons is $E_{\rm kin}=$ 1.61 eV. Figures \ref{fig:Au}(a) and (c) show the $\zeta_z^\pm$ results for the ($M$)-[7]H/Au and ($P$)-[7]H/Au interfaces, respectively, at the $\Gamma$ point. Similarly, Figures \ref{fig:Au}(d) and (f) show the $\zeta_z^{\rm LP}$ results for the two interfaces. Additionally, $\zeta_x^\pm$ and $\zeta_y^\pm$ results at $\Gamma$ are presented in Figure S4. $\zeta_x^\pm$, $\zeta_y^\pm$, and $\zeta_z^\pm$ results at another $\mathbf{k}_\parallel$ point away from $\Gamma$ are presented in Figure S5. Compared to the clean Au (111) surface, the adsorption of the chiral molecules substantially modifies the energy-resolved spin polarization. In the presence of molecular adsorbates, Kramers degeneracy remains at $\Gamma$ while spin splitting appears away from $\Gamma$. 

To determine whether these changes are genuinely enantiomer specific, we analyze the symmetry relation between the two helicene/Au interfaces. After a careful symmetry analysis of the data for all $\mathbf{k}_\parallel$ points considered in the Brillouin zone, we find that the following relations hold:
\begin{subequations}
\begin{align}
\zeta_x^\pm(M;k_x,k_y) & = -\zeta_x^\mp(P;k_x,-k_y); \\
\zeta_y^\pm(M;k_x,k_y) & = \zeta_y^\mp(P;k_x,-k_y); \\
\zeta_z^\pm(M;k_x,k_y) & = -\zeta_z^\mp(P;k_x,-k_y).
\end{align}
\label{eq:sym}
\end{subequations}
These relations arise because the ($M$)-[7]H/Au and ($P$)-[7]H/Au systems are related by a mirror reflection with respect to the $xz$ plane, which maps $k_y \to -k_y$ and transforms the spin polarizations accordingly. In addition, the definitions of the $x$ and $y$ directions (see Figure S1 of the Supporting Information), as well as the molecular orientations on the surface, must be taken into account when interpreting the in-plane spin polarization. In contrast, the out-of-plane component of the spin polarization is the most straightforward quantity for the present comparison and is also the quantity often measured in experiment. In particular, at the $\Gamma$ point (corresponding to normal emission in PES), symmetry requires that $\zeta_z^\pm(M)=-\zeta_z^\mp(P)$. We emphasize that the symmetry relations in Eq. \eqref{eq:sym} are not, by themselves, evidence that the interface enhances or suppresses PES spin polarization. Rather, they establish the baseline expectation for two enantiomeric interface structures that are exact mirror images of one another. Thus, within the present theoretical framework, the two enantiomers are not expected to produce independent normal-emission responses; rather, their $\Gamma$-point signals are symmetry linked. Any enantiomer-dependent asymmetry in $\zeta_z$ can only appear away from $\Gamma$. 

Notably, for an incident light that is linearly polarized, Eq. \eqref{eq:sym} reduces to $\zeta_z^{\rm LP}(M;\Gamma)=-\zeta_z^{\rm LP}(P;\Gamma)$. As can be seen in Figures \ref{fig:Au}(d)(f), although these values are much smaller than their counterparts for circularly polarized incident light, the signs of $\zeta_z^{\rm LP}$ are indeed opposite for the two enantiomers. This sign reversal superficially resembles the handedness-dependent spin polarization often associated with CISS.\cite{Gohler2011,Kettner2018} However, in the present calculations it should not be interpreted by itself as evidence for a molecular spin-filtering/polarizing mechanism, because the sign reversal follows from the symmetry relation between the two mirror-related interfaces under optical excitation with linearly polarized light.

This symmetry analysis clarifies the main trend in Figure \ref{fig:Au}. Although the two enantiomers -- ($M$)-[7]H and ($P$)-[7]H -- modify the spin polarization of photoelectrons emitted from the Au (111) surface, their effects are qualitatively similar and symmetry related. In other words, within the present first-principles treatment, the dominant effect of adsorption on Au (111) is to reshape the interfacial photoemission response, rather than generating a large, systematic, enantiomer-dependent shift in the out-of-plane spin polarization, especially for circularly polarized incident light. This conclusion is qualitatively consistent with Ref. \citenum{baljozovic2023} and is less supportive of interpretations that attribute every adsorbate-induced change in PES spin polarization directly to chirality alone. Within this framework, the apparent enantiomer-dependent signs of $\zeta_z^{\rm LP}$ for linearly polarized incident light can be better understood: because $\zeta_z^{\rm LP}(\Gamma)=0$ for clean Au (111), the qualitatively similar molecular modulation of the interfacial electronic structure results in non-zero spin polarizations but with opposite signs for the two symmetry-related enantiomers, i.e., $\zeta_z^{\rm LP}(M;\Gamma)=-\zeta_z^{\rm LP}(P;\Gamma)$.


\begin{figure}[!ht]
\centering
\includegraphics[width=3.3in]{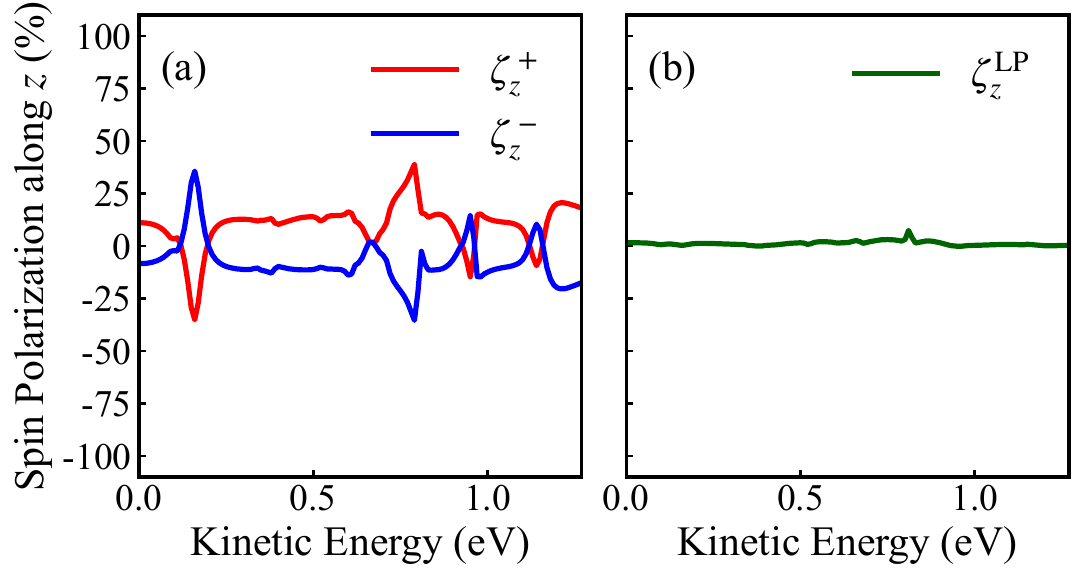}
\caption{Spin polarization along the $z$ direction, for coronene adsorbed on Au (111), at the $\Gamma$ point. (a) Incident light is circularly polarized. Red (blue) lines denote $\zeta^+$ ($\zeta^-$). (b) Incident light is linearly polarized.}
\label{fig:coro}
\end{figure}

The key question is whether the changes found for helicene/Au interfaces arise specifically from chirality, or whether similar changes can result more generally from adsorption-induced modification of the interface. To address this question, we examine a non-chiral control system, coronene adsorbed on Au (111), as shown in Figure \ref{fig:stru}(c). This interface has a work function of 4.54 eV, leading to the maximum $E_{\rm kin}$ being 1.29 eV for an incident light with $h\nu=$ 5.83 eV. The corresponding out-of-plane spin polarizations, $\zeta_z^\pm$ and $\zeta_z^{\rm LP}$, are computed at the $\Gamma$ point, as presented in Figure \ref{fig:coro}(a) and (b), respectively. The symmetry of the non-chiral interface enforces $\zeta_z^+=-\zeta_z^-$ and $\zeta_z^{\rm LP}=0$ at $\Gamma$. As seen from these results, coronene induces changes in $\zeta_z^\pm$ that are qualitatively similar to those produced by either ($M$)-[7]H or ($P$)-[7]H, for circularly polarized incident light. This similarity shows that substantial adsorbate-induced changes in PES spin polarization are not necessarily tied to the chirality of the adsorbate. For the same incident light energy, the lowered work function at the molecule-metal interface enables more occupied states to contribute to PES, including hybridized molecule-metal states. Together with the associated changes in the transition matrix elements, this provides a natural explanation for why non-chiral and chiral adsorbates can produce qualitatively similar modifications.

To further probe the role of substrate SOC, we investigate ($M$)-[7]H and ($P$)-[7]H adsorbed on Cu (111), a system examined experimentally in Ref. \citenum{baljozovic2023}. The corresponding spin polarizations, shown in Figure S6 of the Supporting Information, are much smaller than those for Au and average close to zero over the accessible $E_{\rm kin}$ range. They therefore reinforce the importance of large substrate SOC and indicate that molecules alone cannot be viewed as the sole source of the photoemission spin polarization, at least within the present theoretical framework.

\begin{figure}[!ht]
\centering
\includegraphics[width=3.3in]{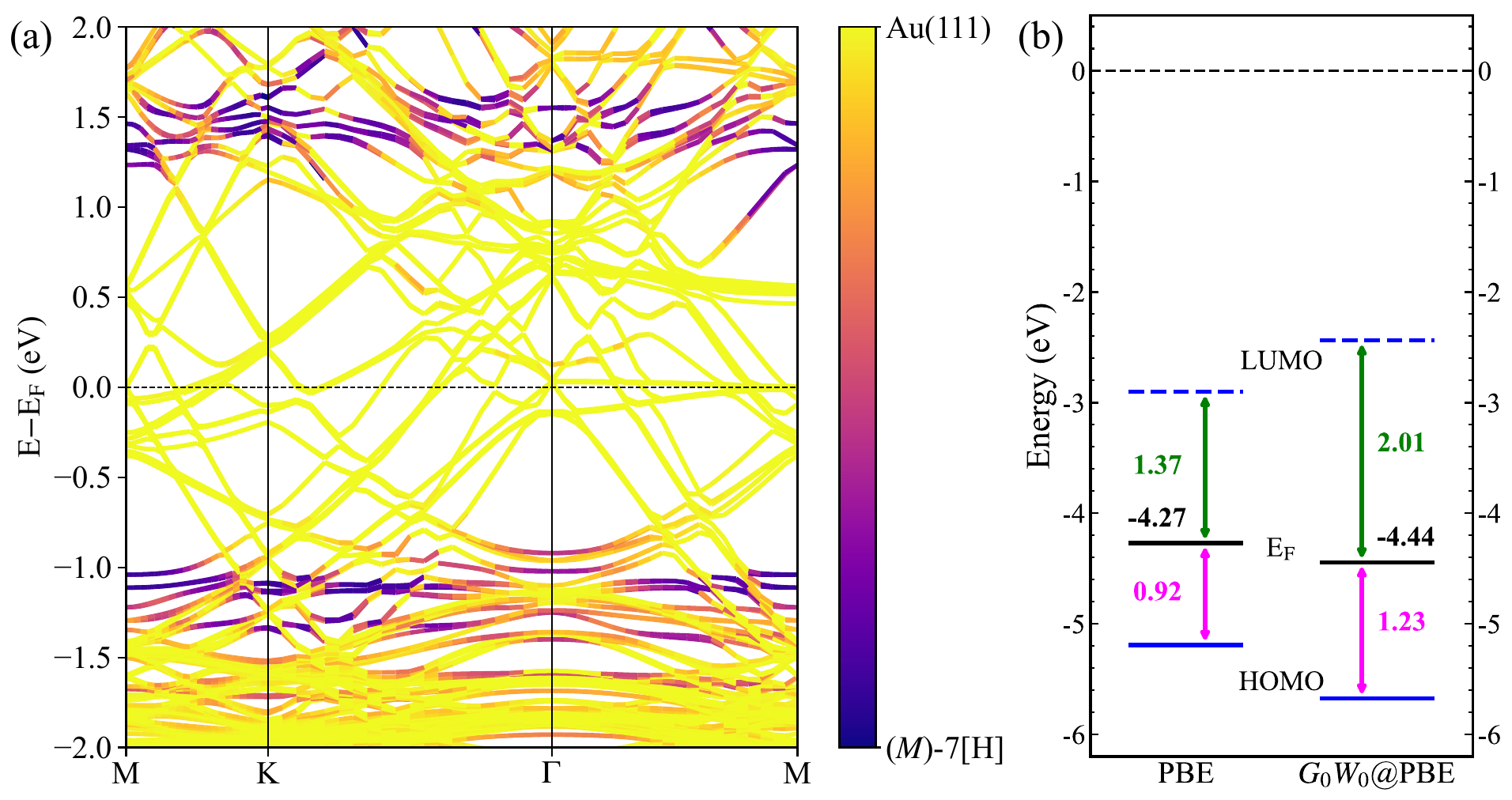}
\caption{Electronic structure of the ($M$)-[7]H/Au interface that includes four layers of Au (111). (a) PBE band structure, with color denoting the projected character of each orbital. The Fermi level of the interface is set to be zero. (b) PBE and $G_0W_0$@PBE energy level alignment at the $\Gamma$ point. Solid and dashed blue lines represent the highest occupied molecular orbital (HOMO) and the lowest unoccupied molecular orbital (LUMO) of the ($M$)-[7]H, respectively. Black lines represent the Fermi level of the interface. Vacuum level is set to be zero.}
\label{fig:es}
\end{figure}

All results presented thus far are based on DFT electronic structures. Here, we briefly assess the impact of many-body corrections, focusing on improved level alignment at the molecule-metal interface.\cite{liu2025} Figure \ref{fig:es}(a) shows the band structure of the ($M$)-[7]H/Au interface, with bands color-coded by their projected character. This calculation employs the Perdew-Burke-Ernzerhof (PBE) functional\cite{PBE} with SOC and includes four Au (111) layers to reduce the computational cost. At the $\Gamma$ point, we observe appreciable hybridization between molecular orbitals and Au states. In particular, the projected HOMO of ($M$)-[7]H lies approximately 0.92 eV below the Fermi level. The reduced work function at the four-layer interface model implies that incident light with $h\nu=$ 5.83 eV probes states up to 1.56 eV below the Fermi level, thereby encompassing these hybridized states. To connect the electronic structure to the $\zeta_z(E_{\rm kin})$ results in Figure \ref{fig:Au}, the initial states contributing to low-$E_{\rm kin}$ features in $\zeta_z$ have stronger molecular character, whereas those contributing to high-$E_{\rm kin}$ features have stronger metallic character. We emphasize, however, that the calculated PES spin polarization also depends on the final states, transition matrix elements, and momentum filtering in Eq. \eqref{eq:isz}, so this projection analysis should be interpreted as a qualitative assignment rather than a one-to-one decomposition of the PES signal.

Given the well-known tendency of PBE to underestimate interfacial level alignment,\cite{neaton2006} we further perform first-principles $G_0W_0$@PBE calculations\cite{HL86,BGW} using the substrate screening approximation\cite{subs-screening} to obtain a more accurate electronic structure, shown in Figure \ref{fig:es}(b), with computational details provided in the Supporting Information. At the $GW$ level, the HOMO is shifted to 1.23 eV below the Fermi level, but remains within the energy window accessible in PES. We also note that the Au $d$ states, located around 2 eV below the Fermi level, lie outside this energy window and therefore do not contribute to the photoemission signal considered here. This supports the conclusion that corrections in the level alignment alone do not overturn the qualitative interface-based interpretation developed from the DFT results.

Several limitations of the present approach should be kept in mind. Our treatment focuses on the first step (optical excitation) of the three-step photoemission model and employs an independent-particle description of the optical excitation. This framework is useful for identifying qualitative trends controlled by the hybridized interface, but it does not yet include all effects needed for a quantitatively complete comparison with experiment. In particular, we do not include the matrix-element renormalization beyond the independent-particle picture, the more realistic final-state description used in one-step photoemission theory\cite{braun1996} (treating the final state as a time-reversed low-energy electron diffraction state), electron-vibrational coupling, or structural order/disorder. These factors may be important when making a detailed comparison with experiment, especially because the spin polarization strongly depends on the chosen spin-analysis axis (in-plane vs. out-of-plane), kinetic energy of photoelectrons, crystal momentum, polarization of incident light, and the set of initial states contributing to the signal (or equivalently, the energy of incident light). From an experimental perspective, spin polarization might be sensitive to the acceptance angle, sample quality, atomic-scale interface structure, and precise data-processing procedure. 

In summary, we developed a first-principles framework for spin polarization in photoemission from molecule-metal interfaces and applied it to helicene adsorbed on Au (111) and Cu (111), with coronene/Au as a non-chiral control. Within the present theory, the conclusions are: (i) molecular adsorption can strongly modify the spin polarization of photoelectrons, compared to that of the clean metal surface; and (ii) this modification appears to be primarily an interface effect rather than systematically enantiomer-dependent, consistent with Ref. \citenum{baljozovic2023}. In other words, this modification is due to the molecular modulation of the hybrid interfacial electronic structure, which is qualitatively similar for the two enantiomers and even for a non-chiral control molecule. The observed relationships between the spin polarizations of the two enantiomers then follow from symmetry requirements. Therefore, the molecular adsorption effect needs to be carefully disentangled from the chirality-dependent effects. A more refined interpretation of the experimental observables is also needed, because the measured change in the spin polarization after molecular adsorption should not automatically be attributed to molecular chirality alone.

\section{Supporting Information Description}
Theoretical details in computing photoemission; Orientations of the real- and reciprocal-space lattice vectors and the 2D Brillouin zone; spin polarization of Au (111) unit cell at different representative $\mathbf{k}_\parallel$ points; spin polarization of Au (111) unit cell along the $z$ direction using different energies of incoming light; additional results for the spin polarization of ($M$)-[7]H/Au and ($P$)-[7]H/Au interfaces; spin polarization of ($M$)-[7]H/Cu and ($P$)-[7]H/Cu interfaces; computational details of geometry optimization; computational details of $G_0W_0$ calculations.


\section{Acknowledgments}
We thank Thomas Frederiksen, Per Hedeg\r{a}rd, Jeff Neaton, Latha Venkataraman, and Binghai Yan for fruitful discussions. We also thank an anonymous reviewer for suggesting the discussion of linearly polarized incident light. This work is supported by an NSF CAREER award, DMR-2044552 and used Bridges-2 at the Pittsburgh Supercomputing Center through Allocation No. PHY220043 from the Advanced Cyberinfrastructure Coordination Ecosystem: Services \& Support (ACCESS) program, which is supported by NSF Grants No. 2138259, No. 2138286, No. 2138307, No. 2137603, and No. 2138296. Very large-scale calculations were carried out using resources of the National Energy Research Scientific Computing Center, a DOE Office of Science User Facility supported by the Office of Science of the U.S. Department of Energy under Contract No. DE-AC02-05CH11231 using NERSC award BES-ERCAP0036533. A.A. acknowledges a Rumble Fellowship and an A. Paul and Carole C. Schaap Endowed Distinguished Graduate Award from the Graduate School at Wayne State University. Z.-F.L. acknowledges an Alfred P. Sloan Research Fellowship, No. FG-2024-21750.
\bibliography{lib}

\begin{tocentry}
\center
\includegraphics[scale=0.3]{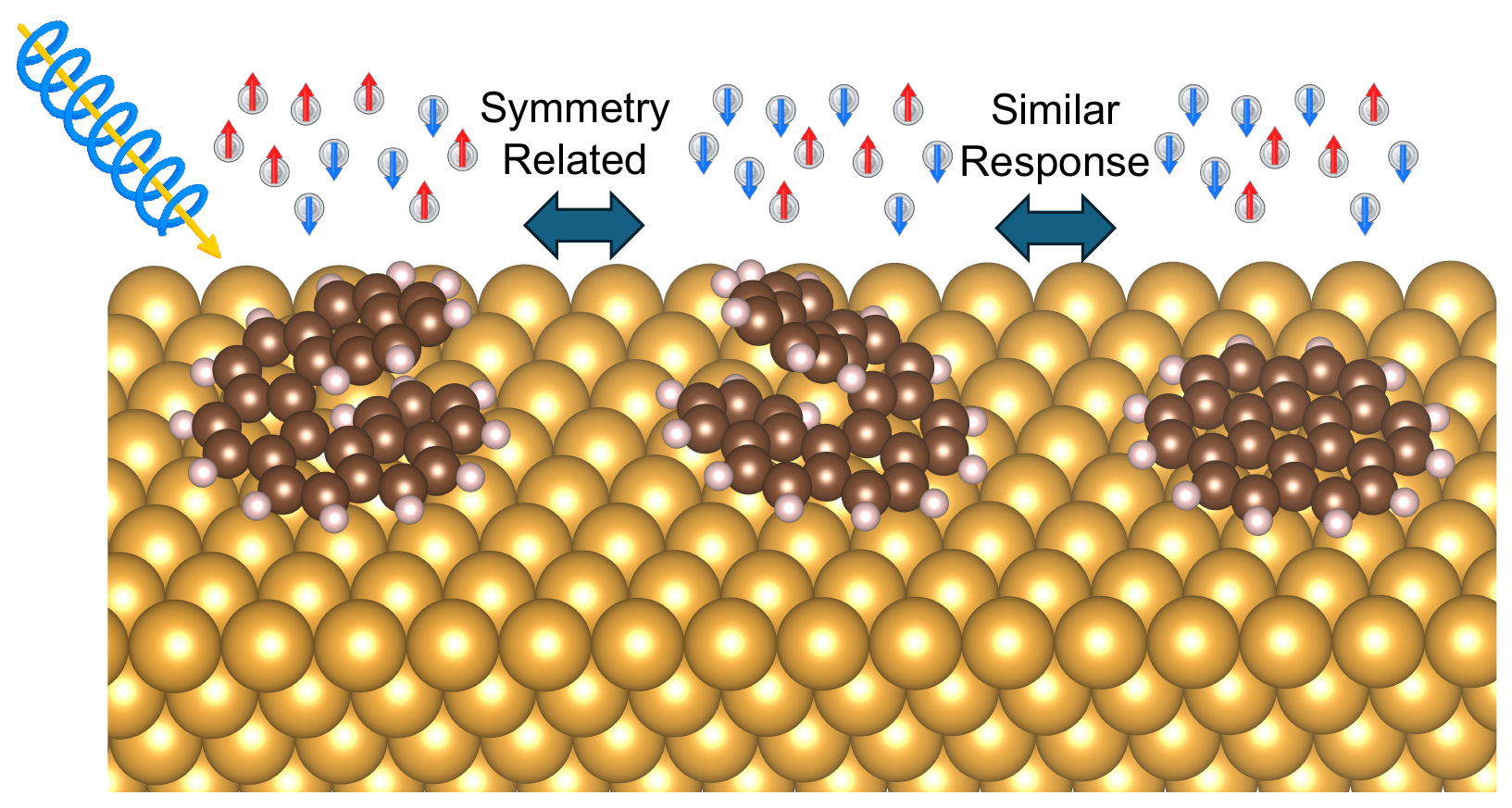}
\end{tocentry}

\end{document}